\documentclass{aastex62}
\usepackage{amssymb}

\received{October 29, 2018}
\revised{\today}
\accepted{...}
\submitjournal{ApJ}
\shorttitle{radio emission in type III solar radio bursts}
\shortauthors{Krasnoselskikh et al}

\begin{document}

\title{On the mechanism of radio emission in type III solar radio bursts}

\correspondingauthor{Vladimir Krasnoselskikh}
\email{vkrasnos@cnrs-orleans.fr}

\author{Vladimir Krasnoselskikh}
\affiliation{LPC2E/CNRS, UMR 7328, 3A Avenue de la Recherche Scientifique, Orleans, France}

\author{Andrii Voshchepynets}
\affiliation{The Swedish Institute of Space Physics (IRF), Rymdcampus 1, Kiruna, Sweden}

\author{Milan Maksimovic}
\affiliation{ LESIA/CNRS, Paris observatory, France}

\begin{abstract}
Type III solar radio bursts are generated by streams of energetic electrons
accelerated at the Sun during periods of the solar activity. The generation
occurs in two steps. Initially, electron beams generate electrostatic
Langmuir waves and then these waves are transformed in electromagnetic
emissions. It is widely accepted that the mechanism of generation of
emission on fundamental frequency close to plasma frequency is due to
induced scattering of Langmuir waves into electromagnetic. However this
process imposes quite restrictive limit of the ratios of effective
brightness temperatures of electromagnetic and Langmuir waves in the source
region. Recent studies showed that the level of density fluctuations in the
solar wind and in the solar corona is so high that it may significantly
affect beam-plasma interaction. Here we show that the presence of intense
density fluctuations not only crucially influence the process of beam plasma
interaction but also changes the mechanism of energy transfer from
electrostatic waves into electromagnetic. Reflection of the Langmuir waves
from the density inhomogeneities may result in partial transformation of the
energy of electrostatic wave into electromagnetic. We show that the linear
wave energy transformation for the level of fluctuations of the order of 1\%
or higher may be significantly more efficient for generation of type III
solar radio bursts than conventionally considered process of nonlinear
conversion due to induced scattering on ions.
\end{abstract}

\keywords{acceleration of particles--- Sun: radio radiation--- Sun: particle
emission--- solar wind}

\section{Introduction}

Solar type III radio bursts are amongst the strongest radio emissions in the
heliosphere. It is wildly accepted that the high energy electrons $\sim 5-30$%
~keV, accelerated during reconnection of the magnetic flied lines in solar
atmosphere, are responsible for generation of these radio emission %
\citep{1985srph.book..289S,2017RvMPP...1....5M}. Important characteristic of
the type III radio bursts is the fast frequency drift rate %
\citep{Wild_McCready_1950}. Type III bursts can start at a frequency of
several hundred of MHz and then go down to tens kHz within few minutes with
the increasing duration at a given frequency\citep
{1973SoPh...30..175A,Reid_Kontar_2018,2018ApJ...857...82K}. To explain the
frequency drift, the beams should have near relativistic speeds $0.1-0.5c$ %
\citep{1990SoPh..130....3M} and generate radio emission near plasma
frequency, $f_{pe}$ and double plasma frequency (harmonic) %
\citep{Ginzburg_Zhelezniakov_1958}. Frequency of the emission follows local
electron density along the beam path, starting at high frequencies in dense
plasma close to the Sun and then decrease over time as the beam propagates
in the expanding solar corona and later solar wind \citep{Krupar_et_al_2018}.

Generation of radio waves occurs in two steps. In the original study, %
\citet{Ginzburg_Zhelezniakov_1958} proposed that the two-stream instability
of an electron beam results in the growth of electrostatic Langmuir waves
(ES) that later produce electromagnetic (EM) emission at the plasma
frequency due to the scattering on ions while the coalescence of two
Langmuir waves can produce the harmonic emission. The theory has been
subsequently refined and alternative mechanisms for the conversion of the
beam-driven Langmuir waves into electromagnetic radiation have been proposed %
\citep{1990SoPh..130....3M,Malaspina_et_al_2012}. While exact mechanism of
the ES to EM conversion is still under debate, the generation of the
Langmuir waves by electron beams in the solar wind has been confirmed by 
\textit{in situ} measurements \citep{Lin_et_al_1981,Ergun_et_al_1998}.

Beam-type configurations in a plasma are known to be unstable and relaxation
of the beam-plasma system to the stable state results in growth of the
Langmuir turbulence %
\citep{Romanov_Filippov_1961,Drummond_Rosenbluth_1962,Vedenov_1963}. Landau
resonance enables effective energy transfer from the beam electrons to the
waves, as a result up to $2/3$ of the initial beam energy can be transferred
to the Langmuir waves through wave-particle interaction %
\citep{Vedenov_Ryutov_1975}. Recent studies showed that the level of density
fluctuations in the solar wind and in the solar corona is so high that they
may significantly affect beam-plasma interaction %
\citep{2001A&A...375..629K,Zaslavsky_et_al_2010,Krafft_Volokitin_Krasnoselskikh_2013,Reid_Kontar_2013,Voshchepynets_et_al_2015}%
. The effect of density fluctuations results in phase velocity variations of
Langmuir waves. These variations change the wave resonance velocities of the
electrons. This leads to significant decrease of the increment of
instability and important increase of the relaxation length of the beam %
\citep{Voshchepynets_Krasnoselskikh_2015}.

Presence of the density fluctuations has significant impact on the observed
properties of the Langmuir waves (see \citet{Krasnoselskikh_2007} and
references therein). First, this idea has been proposed in %
\citet{Smith_Sime_1979} in order to explain observed clumping of Langmuir
waves in type III source regions. Later, it has been proposed %
\citep{Ergun_2008} that the density irregularities in the solar wind can
take form of cavities that might result in modulation of the the waveforms
of the Langmuir waves. Analysis of the large number of individual waveforms
measured by STEREO and WIND showed good agreement with theoretical
predictions \citep{Ergun_2008,Malaspina_Ergun} and results of the numerical
simulations \citep{Krafft_2014}. To address stochastic nature of the density
fluctuations, several statistical models that deduced properties of the
Langmuir waves from the probability distribution function of the density
fluctuations have been proposed. Stochastic growth theory proposed by %
\citet{Robinson_1995} predicted that distribution of the amplitudes of the
Langmuir waves in type III source regions should follow log-normal
distribution. While some observations %
\citep{Robinson_Cairns_Gurnett_1993,Piza_2015} show good agreement with
these predictions, there are numerous reports of observations and
simulations %
\citep{Krasnoselskikh_2007,Vidojevic_2012,Reid_Kontar_2017,Voshchepynets_2017}
that deviations of the distribution of the amplitudes of the Langmuir waves
from log-normal can be rather significant.

The aim of the present study is to determine the role of the density
fluctuation in the conversion process of the beam-generated ES waves into EM
emissions. Reflection of the Langmuir waves from the density inhomogeneities
may result in partial transformation of the energy of electrostatic wave
into electromagnetic. We consider this effect of linear wave energy
transformation in application to generation of type III solar radio bursts.
We use the probability distribution of density fluctuations to evaluate the
statistical characteristics of such process and its efficiency. We show that
the mechanism of linear transformation for the relative density fluctuations
of the order of $1\%$ may be significantly more efficient than widely
accepted process of nonlinear conversion of Langmuir waves due to induced
scattering on ions.

\section{Calculation of the efficiency of energy conversion}

When the density fluctuations along the waves path cannot be neglected, the
propagation of the wave can be described with non-linear Bohm-Gross
dispersion relation for Langmuir waves : 
\begin{equation}
\omega ^{2}=\omega _{p}^{2}\left( 1+3\lambda _{D}^{2}k^{2}+\frac{\delta n}{N}
\right),
\end{equation}
where $\omega $ and $k$ are frequency and wave-vector of the Langmuir wave, $%
\omega _{p}$ is the electron plasma frequency for the electron number
density $N$, $\delta n$ is a deviation of the density from the average value 
$N$, and $\lambda _{D}$ is the Debye length. The waves are assumed to be
generated resonantly: $\omega =kV_{b}$, where $V_{b}$ is the beam velocity
that significantly exceeds thermal velocity of electrons $v_{T}=\sqrt{T/m}$ (%
$T,m$ are electron temperature and mass). A Langmuir wave propagating in
plasma with density inhomogeneities encounters density depletions and
enhancements along its path. When the wave goes to the increasing density
region where the local plasma frequency $\omega _{p}$ becomes equal to the
wave frequency $\omega $, the wave is reflected.

\begin{figure}[ht!]
\includegraphics[width=0.99\linewidth]{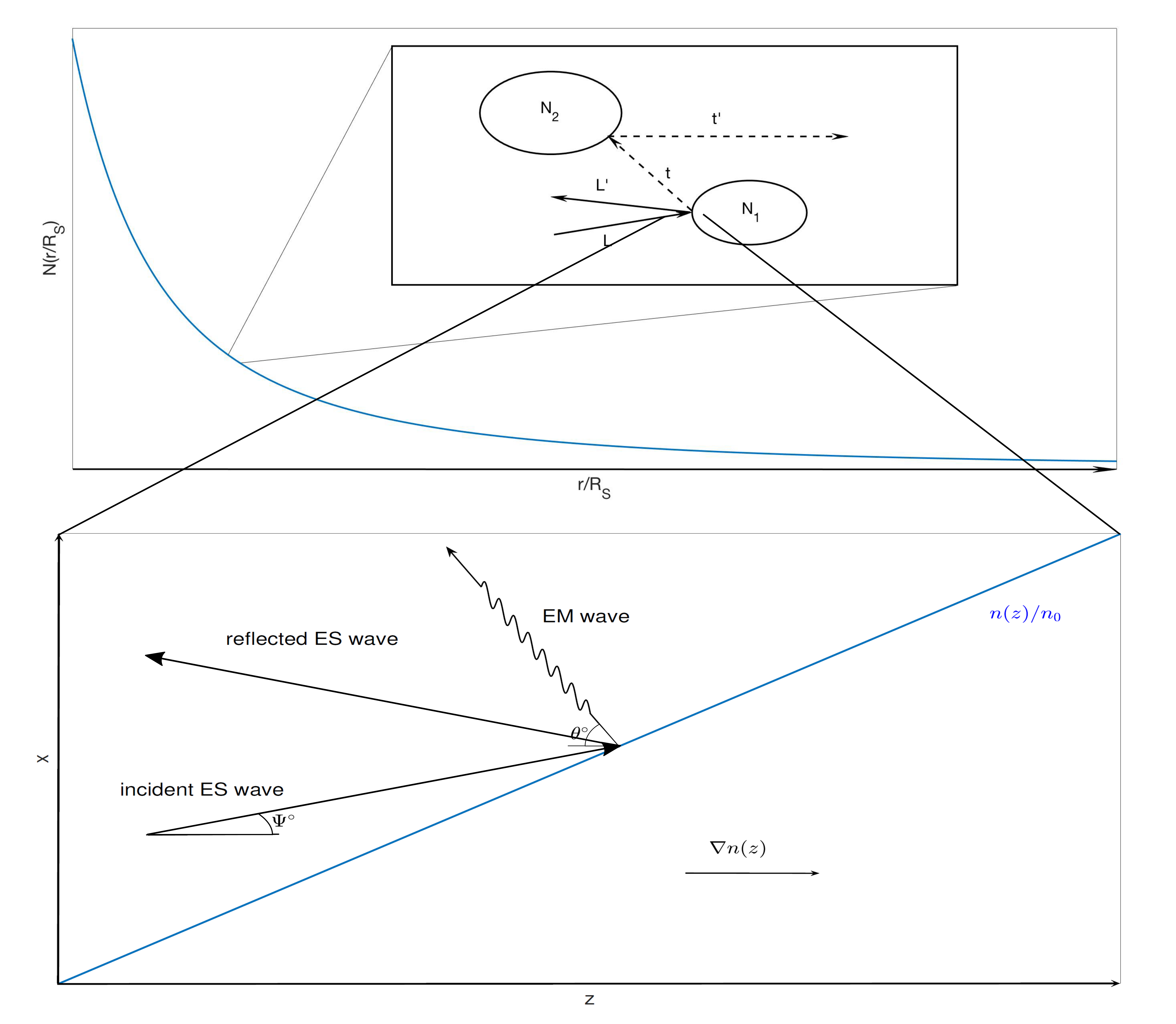}
\caption{Schematic illustration of the formation of the type III solar radio
bursts.}
\label{fig_1}
\end{figure}

Assuming that the density fluctuations are isotropic, the incident angles of
Langmuir waves are distributed uniformly over a semi-sphere. For the
majority of incidence angles, the reflection resembles mirror type
reflection: the $k$-vector component parallel to the direction of the
density gradient changes its direction, while the $k$-vector component
perpendicular to the gradient remains unchanged. In the rather narrow range
of incidence angles, electrostatic waves may couple with the electromagnetic
and the process becomes a three-wave coupling process. In this case, the
reflection results in a Langmuir wave and an electromagnetic wave, so that a
part of the incident Langmuir wave energy is transformed to the
electromagnetic wave. It is worth noting that initial reflection generates
electromagnetic wave propagating in the direction towards the Sun. However,
as an average density decreases with the distance from the Sun the secondary
reflections that may be considered as mirror type will turn the wave
direction towards the Earth (top panel in Figure \ref{fig_1}).

\begin{figure}[th]
\includegraphics[width=0.95\linewidth]{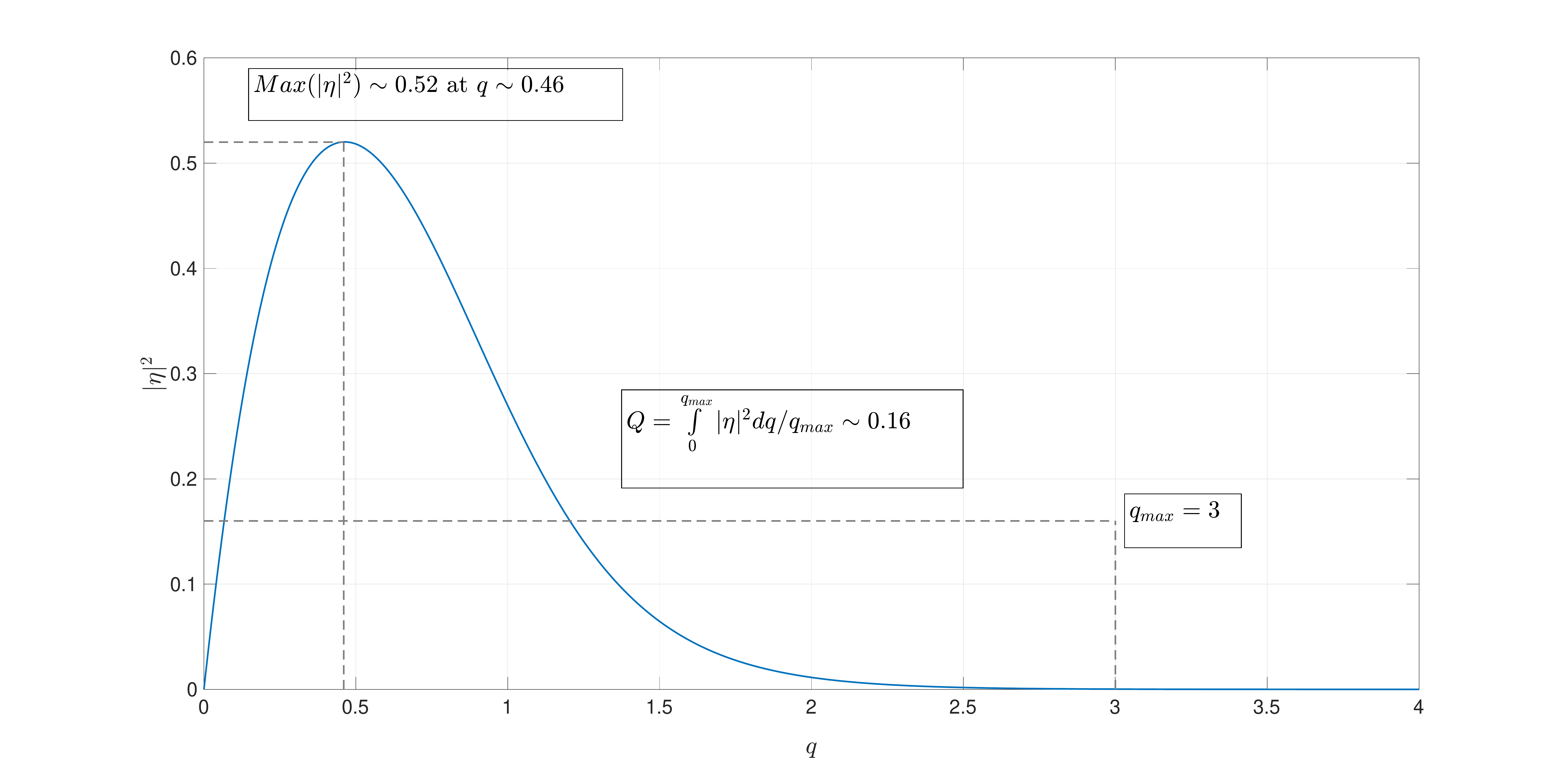}
\caption{Energy transformation coefficient $|\protect\eta |^{2}$ of the
incident Langmuir wave to electromagnetic wave as a function of parameter $%
q=[c/(\protect\omega L)]^{2/3}\sin ^{2}\protect\theta $. }
\label{fig_tr_coef}
\end{figure}

Following the previous works \citep{Piliya_1966,Stenzel_1974}, let the
Langmuir wave of frequency $\omega $ propagates obliquely to the direction
of the density gradient that we choose to be along the $z$-axes (bottom
panel in Figure \ref{fig_1}). Let the perpendicular component of the $k$%
-vector be directed along the $x$-axes and be equal to $k_{x}=(\omega
/c)\sin \theta $, so the component along the $z$-axes is 
\begin{equation}
k_{z}=\sqrt{\frac{\omega ^{2}-\omega _{p}^{2}}{(3T/m)}-k_{x}^{2}},
\end{equation}
The angle $\theta $ corresponds to the angle of propagation of the reflected
electromagnetic wave with respect to $z$-direction. The incidence angle of
electrostatic wave $\psi $ is determined by the ratio of perpendicular and
parallel components of ${k}$-vector, i.e. $\tan \psi =(V_{b}/c)\sin \theta$.
When the density profile is linear function of distance with characteristic
scale $L$, one finds 
\begin{equation}
\frac{N+\delta n(z)}{N}=\frac{\omega _{p}^{2}(z)}{\omega _{p0}^{2}}=\left(
1+ \frac{z}{L}\right),
\end{equation}
so the incident Langmuir wave reflects when the local plasma frequency is 
\begin{equation}
\omega _{p}(z)=\omega \sqrt{1-\frac{3T}{mc^{2}}\sin ^{2}\theta }.
\end{equation}
where $c$ is the speed of light. A fraction of incident wave energy is
reflected as Langmuir wave, while the other part is converted to an
electromagnetic wave in this mode conversion point $\omega _{p}(z)=\omega $.
The electromagnetic wave propagates outward into the direction of the
density decrease beyond its cutoff frequency at $\omega _{p}(z)=\omega \cos
\theta $. The problem has been studied by many authors starting with the
pioneering work by \citet{Denisov_1957}. Several methods have been developed
to evaluate the conversion coefficient, \citep[e.g. review by
][]{Piliya_1966}. Recently, \citet{HL_1,HL_2} have performed analytical
study and obtained an analytical expression for the conversion coefficient: 
\begin{equation}
\mid \eta \mid ^{2}=1-\mid R\mid ^{2}=\frac{(2\pi )^{2}q[Ai^{\prime
}(q)]^{2} }{[1+\pi ^{2}q(Ai^{\prime }(q))^{2}]^{2}+\pi ^{4}q^{2}[(Ai^{\prime
}(q)]^{2}[Bi^{\prime }(q)]^{2}},
\end{equation}
where $R$ is the reflection coefficient defined as the ratio of the
reflected Langmuir wave amplitude to the amplitude of the incident wave, $%
\eta $ is the ratio of the electromagnetic and the incident Langmuir wave
amplitudes, $q=[c/(\omega L)]^{2/3}\sin ^{2}\theta $, $Ai(q)$ and $Bi(q)$
are Airy functions, and $Ai^{\prime }(q),$ and $Bi^{\prime }(q)$ their
derivatives. The dependence of the energy conversion coefficient on
parameter $q$ is shown in Figure \ref{fig_tr_coef}.

It is convenient to re-write the reflection coefficient in terms of
incidence angle of Langmuir wave $\Psi $. The parameter $q$ can be written 
\begin{equation}
q(L,\Psi )=\left( \frac{c}{\omega L}\right) ^{2/3}\left( \frac{k^{2}c^{2}}{%
\omega ^{2}}\right) \sin ^{2}\Psi ,
\end{equation}%
where $L$ is the characteristic density inhomogeneity scale. Here we take
into account that the conversion coefficient has non-zero values only in the
range of $q,$ $0<q<q_{\max }$. In our probabilistic model, the efficiency of
the beam-generated Langmuir waves conversion into EM emission is evaluated
averaging over angles and the density fluctuation scales. To evaluate the
ensemble averaged values taking into account the probability distribution of
the density fluctuations, we choose hereafter the reference frame, where $z$
-axes is directed along the wave vector of the propagating Langmuir wave.
The average probability can be written as the product of probability
distributions in angle $P(\Psi )$ and in scale $P(L)$: 
\begin{equation}
K_{\text{eff}}=\frac{W_{em}}{W_{L}}=P_{\text{ref}}\int\limits_{0<q<q_{\max
}}P(\Psi )P(L)|\eta (L,\Psi )|^{2}\sin \Psi d\Psi dL,
\end{equation}%
where $K_{\text{eff}}$ is the energy conversion coefficient from ES Langmuir
waves to EM, $W_{em}$ is energy of reflected EM wave, $W_{L}$ is the energy
of incident Langmuir wave, and $P_{\text{ref}}$ is the probability of ES wave
reflection. As it was shown in \citet{Voshchepynets_et_al_2015} and %
\citet{Voshchepynets_Krasnoselskikh_2015}, in the plasma with random density
fluctuations $P_{\text{ref}}$ can be calculated by making use of a
probability distribution function of the amplitudes of the density
fluctuations $P(\delta n/N)$.

Since the conversion occurs only when the parameter $q$ is in the range from
about $0$ to $q_{\max }$, this leads to the limited angular range of
reflected electromagnetic waves given by: 
\begin{equation}
0<\sin \Psi <q_{\max }\left( \frac{c}{\omega L}\right) ^{1/3}\left( \frac{%
\omega }{kc}\right),
\end{equation}
that corresponds to values of the perpendicular component of the $k$-vector
of incident Langmuir wave 
\begin{equation}
0<k_{\perp }<q_{\max }\left( \frac{c}{\omega L}\right) ^{1/3}\frac{\omega }{c%
}.
\end{equation}
Taking into account that the $k$-vector of the Langmuir wave is
approximately equal to $k_{L}\simeq \omega /V_{b}$, the conversion may occur
only when the angle $\Psi $ given by 
\begin{equation}
0<\frac{k_{\perp L}}{k_{L}}=\sin \Psi <q_{\max }\left( \frac{c}{\omega L}
\right) ^{1/3}\frac{\omega }{kc}\simeq q_{\max }\left( \frac{c}{\omega L}
\right) ^{1/3}\frac{V_{b}}{c}.
\end{equation}
To simplify the evaluation of the integrals, we shall take the conversion
coefficient to be approximately constant (corresponding to its average value 
$\mid \eta (L,\Psi )\mid ^{2}=Q\simeq 0.16$ in the range of $q$ between zero
and $q_{\max }$. Under this assumption the integration over angles and
scales $L$ may be carried out independently step by step. Assuming $P(\Psi
)=(2\pi )^{-1}\sin \Psi $, the integration over angles results in 
\begin{equation}
\int\limits_{0}^{\Psi _{\max }}P(\Psi )\sin \Psi d\Psi =\frac{1}{2\pi }
\int\limits_{0}^{\Psi _{\max }}\sin \Psi d\Psi =\frac{1}{4\pi }q_{\max
}^{2}\left( \frac{c}{\omega L}\right) ^{2/3}\left( \frac{\omega }{kc}\right)
^{2}\simeq \frac{q_{\max }^{2}}{4\pi }\left( \frac{c}{\omega L}\right)
^{2/3}\left( \frac{V_{b}}{c}\right) ^{2}.
\end{equation}

Then the conversion coefficient may be re-written: 
\begin{equation}
K_{\text{eff}}=0.5\frac{Q}{4\pi }q_{\max }^{2}\left( \frac{\omega }{kc}
\right) ^{2}\left( \frac{c}{\omega }\right) ^{2/3}\int\limits_{0}^{\infty } 
\frac{1}{L^{2/3}}P\left( L\right) dL.  \label{eq_K_eff_nd}
\end{equation}
where $P(L)$ is a probability distribution of the density gradients
(scales). In order to calculate $P(L)$ one should use spatial profiles of
the density fluctuations. In the present study we use synthetic density data
calculated from published density power spectra. \citet{Kellogg_1999}
proposed a procedure based on the inverse Fourier transform that allows to
reconstruct density profiles $n(t)$ from the power spectrum assuming the
phases of waves to be random. It is known from in situ spacecraft
measurements in the solar wind \citep{Celnikier_1987,Chen_2012} that the
density spectrum can be considered as a broken power-low, with different
spectral indices, about $5/3$ for low frequency part and about $1$ for
higher frequency part with the transition at about $0.6Hz$. In order to
transform these profile to the spatial profiles $n(z)$, one can use Taylor
hypothesis assuming that fluctuations are convected with the characteristic
velocity of the solar wind, $V_{SW}\sim 400km/s$. We used power spectrum in the range of frequencies between $10^{-2}Hz$ and $530Hz$. Lower frequency limit defines maximal length of the
synthetic density profile ($100s$ or $4\cdot 10^{4}km$). Highest frequency
limit defines the smallest scale of the density fluctuation presented in
this profile. In the present study this scale is set to be of approximately $750m$ that for the plasma conditions relevant for 1AU is about $50\lambda
_{D}$. After the profiles were generated, normalization procedure was
applied to ensure that $\left\langle n(z)\right\rangle _{z}=N_{0}$ and $%
\sqrt{\left\langle (n(z)-N_{0})^{2}\right\rangle _{z}}=\delta n$ for each of
the density profiles (here brackets denote averaging). We consider different
levels of density fluctuations, $\delta n/N_{0}$, between $0.01$ and $0.1$.
For more details on the procedure we refer to %
\citet{Voshchepynets_Krasnoselskikh_2015}.

\begin{figure}[th]
\includegraphics[width=0.95\linewidth]{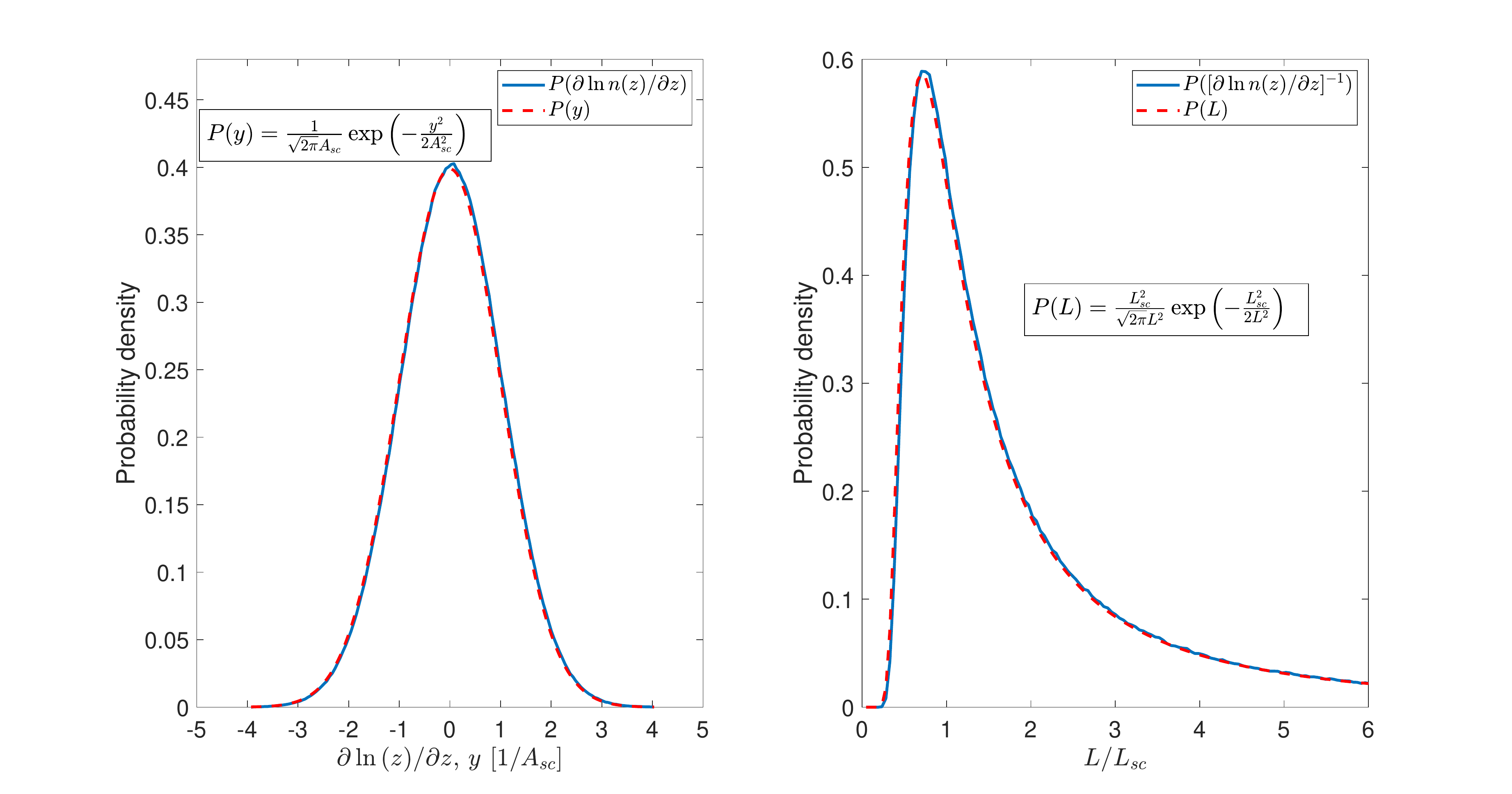}
\caption{Left panel: probability distribution function of the density
fluctuations. Blue line shows distribution of $\ln{\partial n(z)/\partial z}$
obtained from the synthetic profiles of the density fluctuations and red
line shows fit of $P(\ln{\partial n(z)/\partial z})$ by Gaussian function.
Right panel: probability distribution of the density gradients. Blue line
shows distribution obtained from the synthetic profiles and red line show
fit of this distribution by function defined by equation \protect\ref%
{L_dist_eq} }
\label{fig_L_dst}
\end{figure}
Locally the density profiles may be approximated by linear function of $n(z)$%
, thus the probability distribution of the characteristic scales could be
retrieved from the distribution of density gradients, $(1/n)\nabla
n=\partial \ln {n(z)}/\partial z=1/L$. Left panel in Figure \ref{fig_L_dst}
shows normalized probability distribution $P(\partial \ln {n(z)}/\partial z)$
obtained from the large number (about 200) of the density profiles $n(z)$
with level of density fluctuations $\delta n/N_{0}=0.01$. It is found that
the distribution $P(\partial \ln {n(z)}/\partial z)$ is very close to
Gaussian with characteristic scale $A_{sc}$: 
\begin{equation}
P\left( \frac{\partial \ln {(z)}}{\partial z}\right) =P(y)=\frac{1}{\sqrt{
2\pi }A_{sc}}\exp {\left( -\frac{y^{2}}{2A_{sc}^{2}}\right) }.
\end{equation}
Taking into account that $y=1/L$ and $P(L)=P(y^{-1})$ one can get $P(L)$ as
follows: 
\begin{equation}
P(L)=\frac{L_{sc}}{\sqrt{2\pi }L^{2}}\exp {\left( -\frac{L_{sc}^{2}}{2L^{2}}
\right) }  \label{L_dist_eq}
\end{equation}
where $L_{sc}=1/A_{sc}$. The functions $P(y^{-1})$ and $P(L)$ are shown at
left panel in Figure \ref{fig_L_dst}. By making use of $P(L)$ one can
integrate last part in equation for energy conversion coefficient as follows:

\begin{equation}
I=\int \limits_{0}^{\infty} \left(\frac{1}{L}\right)^{2/3}P(L)dL=\frac{%
L_{sc} }{\sqrt{\pi} }\int \limits_{0}^{\infty} \left(\frac{1}{L}%
\right)^{8/3}\exp{\ \left(-\frac{L_{sc}^{2}}{L^{2}}\right)}dL=\frac{%
\Gamma(5/6)}{2\sqrt{\pi} L_{sc}^{2/3}},
\end{equation}
where $\Gamma$ is a Gamma function. Substituting the integral $I$ one can
find $K_{\text{eff}}$: 
\begin{equation}
K_{eff}=0.5Q\frac{q_{max}^{2}}{8\pi\sqrt{\pi}}\left(\frac{V_{b}}{c}
\right)^{2}\left(\frac{c}{\omega L_{sc}}\right)^{2/3}\Gamma(5/6) \simeq 
\frac{Q}{10} \left(\frac{V_{b}}{c}\right)^{2}\left(\frac{c}{\omega L_{sc}}
\right)^{2/3}.
\end{equation}

\begin{figure}[th]
\includegraphics[width=0.8\linewidth]{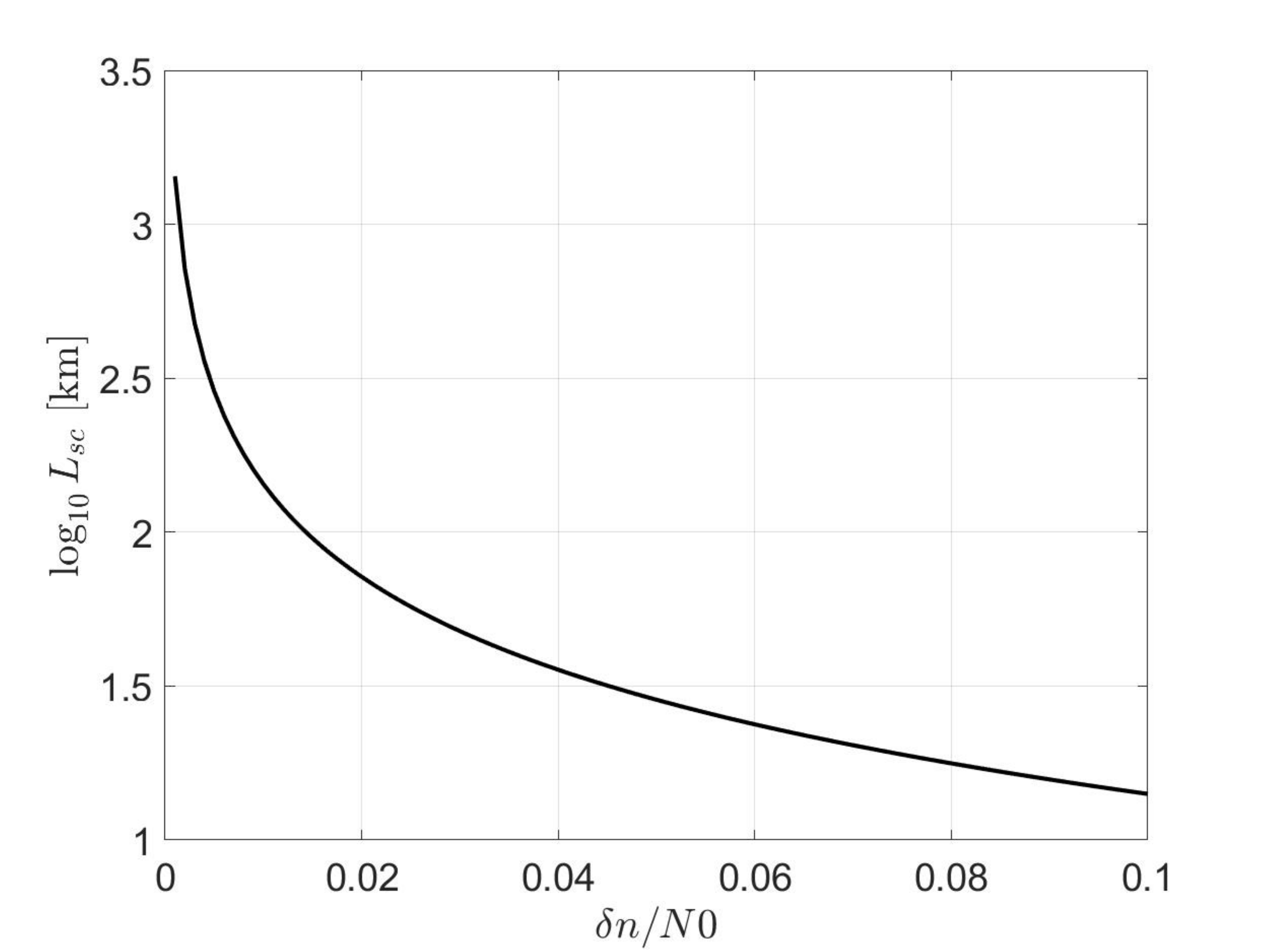}
\caption{Characteristic scale of the density fluctuations $L_{sc}$ as a
function of the level of the density fluctuations $\protect\delta n/N_{0}$.}
\label{fig_L_fnc}
\end{figure}

The characteristic scales of the density gradients obtained from synthetic
density profiles are shown in Figure \ref{fig_L_fnc}. We start with $\delta
n/N_{0}=0.001$ that results in $L_{sc}>10^{3}km$. As one can see $L_{sc}$
drops significantly with increasing level of the density fluctuations, that
in its turn results in increase of the efficiency of energy transformation
from ES to EM waves. Thus for $\delta n/N_{0}=0.1$, that was measured
onboard Helios satellite closer to the Sun \citep{Bavassano_Bruno}
characteristic scale may be less than $15km$.

It is worth noting that the power spectrum used in this study is relevant
for solar wind density fluctuations around 1AU. Closer to the Sun the
spectrum characteristics may be different and as a result density
fluctuations can be described by the different statistics. In order to avoid
speculations (though the method developped here is applicable), we consider
emissions in the frequency range typical for solar type III radio burst
around 1 AU: $10kHz$ to $100kHz$ \citep{Mann1999}.

\begin{figure}[th]
\includegraphics[width=0.95\linewidth]{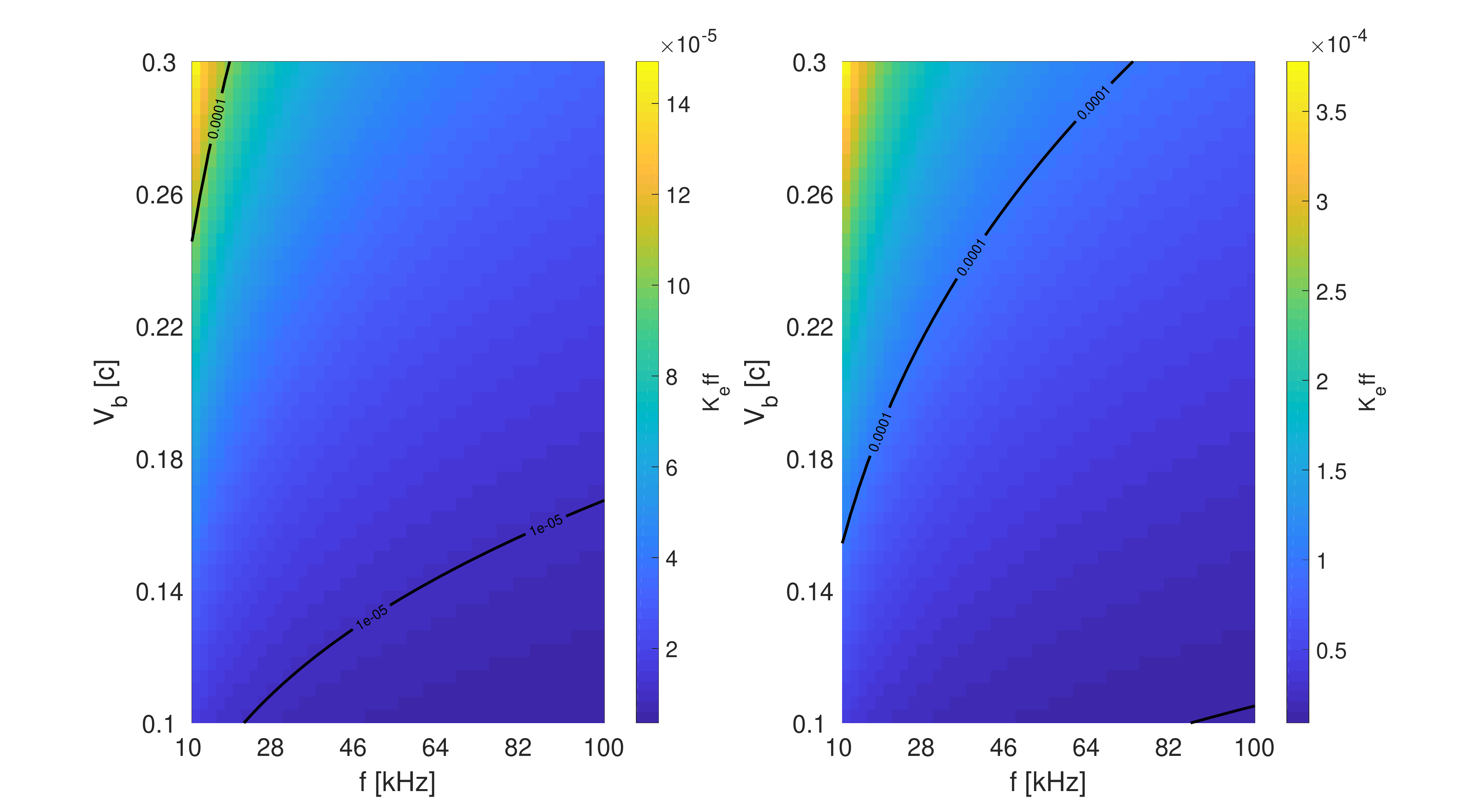}
\caption{Conversion coefficient as a function of beam velocity $V_{b}$ and
frequency of Langmuir waves $f$. Left panel: $\protect\delta n/N_{0}=0.01$.
Right panel: $\protect\delta n/N_{0}=0.04$ }
\label{fig_en_trnfrm}
\end{figure}

Conversion coefficient as a function of beam velocity, $V_{b}$ and Langmuir
wave frequency $f$ is shown in Figure \ref{fig_en_trnfrm}. Left panel shows $%
K_{eff}$ for $\delta n/N_{0}=0,1$. As one can see for $V_{b}>0.15c$
conversion coefficient is above $10^{-5}$ in the whole range of frequencies.
An increase of the level of the density fluctuations results in a decrease
of the characteristic scale of the density gradient. As a result, reflection
of the Langmuir waves will occur more often and $K_{eff}$ will increase.
Right panel in Figure \ref{fig_en_trnfrm} shows $K_{eff}$ for $\delta
n/N_{0}=0.04$. We found that the conversion coefficient is above $10^{-5}$
in the whole frequency range for $V_{b}>0.1c$. For faster beams with $%
V_{b}>0.2c$ conversion coefficient is above $10^{-4}$ for frequencies below $%
50kHz$.

\section{Induced scattering}

It is well known and widely accepted from the very early articles by %
\citet{Ginzburg_Zhelezniakov_1958}, \citet{Kaplan_1969}, \citet{Melrose_1974} and \citet{Melrose_1987} that under condition $T_{e}\sim T_{i}$ (typically in
the solar wind $T_{e}\simeq 2T_{i})$ the damping of ion sound waves is quite
strong and the major nonlinear process of transformation of electrostatic
waves onto electromagnetic (ES-EM) is induced scattering on ions. Generated
electromagnetic waves have frequencies lower than the frequency of the
primary Langmuir waves. To describe this process one should introduce
emission coefficient $U_{\omega ,\Omega }$ for electromagnetic waves that is
defined as the energy per unit frequency interval generated in a unit volume
and in a unit solid angle. This coefficient is expressed making use a power, 
$Q$, radiated by current $j$ excited by Langmuir waves.

For the conversion of plasma waves into electromagnetic due to induced
scattering on ions, $U_{\omega ,\Omega }$ can be found as follows %
\citep{Kaplan_1969}:

\[
U_{\omega ,\Omega }=\frac{k_{t}^{2}}{V_{gr}}\frac{\omega _{p}^{2}}{2(2\pi
)^{5/2}n_{e}}\left( \frac{T_{i}}{T_{i}+T_{e}}\right) ^{2}\int \frac{%
[k_{t}\times k_{L}^{^{\prime }}]^{2}}{k_{t}^{2}k_{L}^{2^{\prime }}}\frac{%
W_{kL}^{^{\prime }}dk_{l}^{^{\prime }}}{V_{ti}|k-k_{L}^{^{\prime }}|}\times
\exp {\left( -\frac{1}{2}\frac{\omega -\omega (k_{L})}{V_{ti}|k-k_{L}|}%
\right) ^{2}} 
\]

For the sake of simplicity we assume $T_{i}=T_{e}$. One can also assume with
the accuracy up to coefficient of the order of unity that radiation is
isotropic (implies that $[k_{t}k_{L}]^{2}=k_{t}^{2}k_{L}^{2}/3$) and $%
k_{L}\gg k_{t}$. This results in:

\[
U_{\omega ,\Omega }=\frac{k_{t}^{2}}{V_{gr}}\frac{\omega _{p}^{2}}{24(2\pi
)^{5/2}n_{e}}\int \frac{W_{kL}^{^{\prime }}dk_{l}^{^{\prime }}}{%
V_{ti}k_{L}^{^{\prime }}}\times \exp {\left( -\frac{1}{2}\frac{\omega
-\omega (k_{L})}{V_{ti}k_{L}}\right) ^{2}} 
\]

Suggesting that for the beam plasma interaction the maximum of the energy
density $W_{kL}$ of the plasma turbulence occurs at a definite value of the
phase velocity $V=V_{b}$, and neglecting the width of the beam choosing $%
W_{kl}=W_{l}\delta (k_{\parallel }-k_{L})$ (here $k_{L}=\omega _{p}/V_{b}$,
and $k_{\parallel }$ is along the direction of the propagation of the beam).
This results in total emission coefficient $U$ that is found integrating
over $\omega $ and $\Omega $:

\[
U=\frac{k_{t}^{2}}{V_{gr}}\frac{\omega _{p}^{2}W_{L}}{48\pi n_{e}} 
\]

The resonance conditions for the wave conversion impose the following
relation between wave vectors of electromagnetic and Langmuir waves: $k_{t}=%
\sqrt{3}k_{L}V_{te}/c$ and $V_{gr}=\sqrt{3}k_{L}V_{te}c/\omega _{p}$. Thus
the one can find:

\[
U=\frac{V_{te}}{V_{b}}\frac{\omega _{p}^{4}}{c^{3}}\frac{\sqrt{3}W_{L}}{%
48\pi n_{e}}. 
\]

One can use total emission coefficient to calculate averaged over period
wave energy density of the EM emission, $W_{t}$. For the comparison with the
linear transformation coefficient determined as  $K_{eff}=W_{t}/W_{l},$ for
ES-EM transformation due to induced scattering on thermal ions it is found
to be equal to:

\[
K_{eff}=\frac{\pi ^{3}}{6\sqrt{3}}\frac{V_{te}}{V_{b}}\frac{V_{te}^{3}}{c^{3}%
}\frac{1}{\lambda _{D}^{3}n_{e}} 
\]

One can see that for typical parameters of the solar wind it is many orders
of magnitude smaller than the  efficiency coefficient for linear
transformation. For the conditions at 1 AU  $T_{e}\sim 10eV,n_{e}=10cm^{-3}$%
, $E_{b}\simeq 10keV$, $K_{eff}\sim 10^{-18}$.

\subsection{Conclusions}

We show that the process of linear conversion of Langmuir waves onto
electromagnetic on density fluctuations can be dominant for the generation
of the type III radio emissions and is significantly more efficient than
conventionally accepted nonlinear process of  induced scattering of Langmuir
waves on ions.

As we show the efficiency of linear conversion is strongly dependent upon
statistical properties of density fluctuations and their gradients. These
characteristics may significantly vary with the distance from the Sun. The
study of these dependencies comes beyond the scope of this paper and will be
addressed in future publications.

\acknowledgments V.K. acknowledges financial support by CNES through grant
"Stereo-Waves invited scientist".

\bibliographystyle{aasjournal}

{}

\end{document}